\magnification1200

\rightline{KCL-MTH-04-03}
\rightline{hep-th/0402140}

\vskip .5cm
\centerline
{\bf  The  IIA, IIB and eleven dimensional theories and their common
$E_{11}$ origin. }
\vskip 1cm
\centerline{Peter West}
\centerline{Department of Mathematics}
\centerline{King's College, London WC2R 2LS, UK}
\vskip 2cm

\leftline{\sl Abstract}
\vskip .2cm
\noindent 
We show that  the commonly considered half BPS solutions of eleven
dimensional supergravity and the ten dimensional type II theories, when
expressed in terms of 
$E_{11}$ group elements, take the universal form 
$\exp(-{1\over 2}ln N \beta\cdot H)\exp((1-N)E_\beta)$.  
Using this formula we find new potential solutions to the $E_{11}$
non-linearly realisations   corresponding to active fields which are
beyond those in the supergravity approximations. These include the space
filling nine brane of the IIB theory. We use $E_{11}$ to give a
correspondence between  the fields of the eleven dimensional  and
the IIA and IIB non-linear realisations without assuming any
dimensional reduction. As one consequence, we find the eleven
dimensional origin of the eight brane solution of the massive IIA theory.

\vskip .5cm

\vfill
\eject
{\bf {1. Introduction }}
\medskip
It is a consequence of supersymmetry that the scalars in supergravity
multiplets belong to  non-linear realisations. The first such example was
in the four dimensional $N=4$ supergravity theory [1] and perhaps the  
 most celebrated example concerns  the four dimensional maximal
supergravity where the scalars belong to a non-linear realisation of 
$E_7$ [2]. A detailed account of the literature on such symmetries  can
be found in the introduction of reference [14]. The eleven dimensional
supergravity theory does not possess any scalars and it was widely
believed that these symmetry algebras were not present in this theory. 
However, it was found that the bosonic sector eleven dimensional
supergravity theory could be formulated as a non-linear realisation [3].
The infinite dimensional algebra involved in this construction was  the
closure of a finite dimensional algebra, denoted
$G_{11}$, with the eleven dimensional conformal algebra. The non-linear
realisation was carried out by ensuring that the equations of motion were
invariant under both finite dimensional algebras, taking into
account that some of their generators were in common.  
The algebra
$G_{11}$ involved the space-time translations together with an algebra
$\hat G_{11}$ 
which contained $A_{10}$ and the Borel subalgebra of $E_7$
as subalgebras. The algebra  $\hat G_{11}$  was not a Kac-Moody algebra,
however,  it was conjectured [4] that the theory could be extended so that
the algebra $\hat G_{11}$  was promoted to  a Kac-Moody algebra.  It was
shown that this Kac-Moody symmetry would have to contain a certain rank
eleven Kac-Moody algebra denoted  $E_{11}$ [4]. 
\par
Consequently, it was argued  [4] that an extension of eleven dimensional
supergravity should possess an $E_{11}$ symmetry that was non-linearly
realised. In particular,  the symmetries found
when  the  eleven dimensional  supergravity theory was dimensionally
reduced would  be  present in this eleven dimensional theory.
One of the advantages of a   non-linear realisation is that the dynamics 
is largely specified by the algebra if the chosen local subalgebra is
sufficiently large. This was not the case for the $G_{11}$ considered
in [3] when taken in isolation as  the
local subalgebra was chosen to be just the Lorentz algebra, but it  is
the case for 
$E_{11}$ with the local subalgebra that was specified in [4].  
The role of space-time and the relation to the later work of [6]
are discussed in the very recent paper [9].  
\par
A similar picture emerged for the IIA and IIB supergravity theories
in ten dimensions and it was conjectured that these theories could be
extended such that they were invariant under $E_{11}$ [4,5]. As a
result, it became clear that   the two different type II theories in
ten dimensions arise by taking different $A_{9}$ subalgebras of
$E_{11}$ to correspond to the gravity sector of the theory. 
[4,5]. 
\par
Arguments  similar  to those advocated for eleven dimensional
supergravity in [4] were proposed to apply to   gravity [12] in D
dimensions and the effective action of the closed bosonic string [4]
generalised to  D dimensions and the underlying Kac-Moody algebras were
identified. It was realised that the algebras that arose in all these
theories were of a special kind and were called very extended Kac-Moody
algebras [13]. Indeed,  for any finite  dimensional semi-simple Lie
algebra
$\cal G$ one can systematically extend its Dynkin diagram by adding three
more nodes to obtain an indefinite 
Kac-Moody algebra denoted $\cal G^{+++}$. In this notation $E_{11}$ is
written as $E_8^{+++}$. The algebras for  gravity and the closed bosonic
string being $A^{+++}_{D-3}$ [12] and $D^{+++}_{D-2}$ [4] respectively. 
\par
It was proposed  in [4,12,13] and [14,15,16,10], 
  that the non-linear realisation of any  very extended algebra 
${\cal G}^{+++}$ leads to a theory, called ${\cal V}_{\cal G}$ in [16],
that at low levels includes gravity and the other fields and it was 
hoped that this non-linear realisation contains an infinite number of
propagating fields that ensures its consistency. 
Indeed, it was shown [16] that the low level content of the adjoint
representation of $\cal G^{+++}$ predicted a field content for a
non-linear realisation of $\cal G^{+++}$ which was in agreement 
with the oxidation theory associated with algebra $\cal G$. 
\par
Some papers have uncovered
relationships between the solutions in the oxidised theories and the
    $\cal G^{+++}$ symmetry conjectured to be present in their
extension. In reference [14],  the non-linear realisation of $\cal
G^{+++}$ restricted to its Cartan subalgebra was constructed and the
resulting Weyl transformations  were shown to transform the moduli of the
Kasner solutions into each  other. Furthermore,  for
$E^{+++}_8=E_{11}$ and $D_{24}^{+++}$ these Weyl transformations  were
shown to be the   U duality
transformations in the corresponding string theories. 
Furthermore in [15]  it was shown that the theories associated with $\cal
G^{+++}$ at low levels, i.e. the corresponding oxidised theories,  admit
BPS intersecting solutions. The results of [14,15] were generalised in
[10]  to an alternative theory;  as in [4] the theory was assumed to be a
non-linear realisation of
$E_{11}$, but the fields were assumed  to depend on an auxiliary
parameter.  The symmetries of the eleven dimensional  theory reduced on
spheres solutions was discussed from the $E_8^{+++}$ perspective in [20]
\par
The aim of the work of [4] was to deduce the underlying
symmetries of M theory in the hope that they would provide a
understanding of what M theory actually is. However, as we will see in
this paper, the presence of such symmetries may be useful to elucidate
some unresolved questions about the relations between the eleven
dimensional theory and the IIA and IIB theories and the branes that occur
in them. As such, we restrict our attention to the case of
$E_8^{+++}$  and  investigate the relationship between the
solutions of eleven dimensional supergravity and IIA and IIB
supergravities and their underlying $E_8^{+++}$ symmetry. One advantage of
a non-linear realisation is that the field content, and in essence  the
dynamics, is determined just by group theory. In particular, the fields
of the non-linear realisation ${\cal V}_{E_8 }$ are contained in the group
element of
$E_8^{+++}$ when modded out by the action of the local subgroup.
Consequently, given any solution we can write down the corresponding
$E_8^{+++}$ group element. In sections three and four  we carry out this
process for the most common half BPS solutions of eleven dimensional
supergravity [32] and the IIA [33] and IIB [34] supergravity theories in
ten dimensional  and find that  the corresponding group element has a
particularly elegant form, namely 
$$g= exp(-{1\over 2}ln N \beta\cdot H)exp((1-N)E_\beta)
\eqno(1.1)$$
where $N$ is a harmonic function and $\beta$ is the  root of  $E_8^{+++}$
whose   corresponding generator is $E_\beta$.   
\par
In section five, we assume that  equation (1.1) also leads to solutions
of the non-linearly realised theory ${\cal V}_{ E_8^{+++}}$ and
derive such solutions for roots which correspond to fields that are beyond
the supergravity theory. In particular, we find the space filling nine
brane of the IIB theory. In section six, we exploit the fact that the
eleven dimensional theory and the IIA and IIB theories have a common 
 $E_8^{+++}$ origin. In particular,   we use this to systematically
 find  relations between the fields and coordinates of the three
theories.  Finally, in section seven we discuss the consequences of this
work. 
\medskip 
{\bf {2.  Kac-Moody algebras and their non-linear realisations}}
\medskip
In this section we recall some of the basic properties of Kac-Moody
algebras [18] and their non-linear realisations. We will
illustrate the general discussion for the case of $E_8^{+++}$. 
\par
A  Kac-Moody  algebras is defined by its Cartan matrix
$A_{ab}$  which by definition 
   satisfies the following properties:  
$$A_{aa}=2, 
\eqno(2.1)$$ 
$$A_{ab}\  {\rm for}\  a\not= b\  {\rm are\  negative\  integers\  or
\ zero}, 
\eqno(2.2)$$ 
and 
$$A_{ab}=0\  {\rm implies}\  A_{ba}=0 .
\eqno(2.3)$$ 
The    Kac-Moody  algebra is  formulated in terms of its
Chevalley generators which consist of the generators
of the  commuting Cartan  subalgebra, denoted  by 
$H_a$, as well as the generators of the positive and negative simple
roots, denoted by 
$E_a$ and  $F_a$ respectively. The Chevalley  generators are taken to
obey  
$$[H_a, H_b]= 0, 
\eqno(2.4)$$
$$[H_a, E_b]= A_{ab} E_b, 
\ \ \ [H_a, F_b]= -A_{ab} F_b, 
\eqno(2.5)$$
$$[E_a, F_b]= \delta_{ab} H_a, 
\eqno(2.6)$$
as well as the Serre relation  
$$[E_a,\ldots [E_a, E_b]\ldots ]= 0, \  
[F_a,\ldots [F_a, F_b]\ldots]= 0
\eqno(2.7)$$ 
In equation (1.7)  there are
$1-A_{ab}$ number of 
$E_a$'s in  the first equation and the same number of $F_a$'s in the
second equation.  Given the generalised Cartan matrix 
$A_{ab}$,  one can uniquely reconstruct the entire
Kac-Moody  algebra  by taking
the multiple  commutators of the simple root generators subject to 
the above  Serre relations.  In particular one can find, at least as a
matter of principle,  the generators and roots of the Kac-Moody
algebra. 
\par
The Cartan matrix is given in terms of the
simple roots $\alpha_a$ by 
$$A_{ab}=2{(\alpha_a,\alpha_b)\over (\alpha_a,\alpha_a)}
\eqno(2.8)$$
In the Cartan-Weyl basis,  the Kac-Moody algebra is generated by  
$H_i$ and $E_a$ and $F_a$
where $H_i$  and $H_a$ are related by 
$H_a=2{\alpha_a^i H_i\over (\alpha_a,\alpha_a)}$ 
where  $\alpha_a^i$ are the components of the simple root $\alpha_a$. 
The  commutator of
$H_i$  with the  generators   $E_a$ and $F_a$ is given by 
$$[H_i, E_a]= \alpha_a^i E_a,\ [H_i, F_a]= -\alpha_a^i E_a .
\eqno(2.9)$$  
It
follows that  a generator $E_\alpha$ associated with root $\alpha$ 
 obeys the commutator
$$ [H_i, E_\alpha]= \alpha^i E_\alpha
\eqno(2.10)$$
and as a result 
$$ [H_c, E_\alpha]= 2  {(\alpha_c,\alpha)\over (\alpha_c,\alpha_c)}
E_\alpha
\eqno(2.11)$$
A result that is obvious with out ever leaving the  original Chevalley
basis by considering 
$E_\alpha$ as a multiple commutator of simple roots and then taking its
commutator with $H_c$. 
\par
Associated with the generator
$E_{\alpha}$ we can associate in a natural way an  element of the
Cartan subalgebra which is  given by 
$$ \alpha^i H_i
\eqno(2.12)$$
It is  the Borel sub-algebra of this $A_1$  which will play
such a central role in this paper.  
\par 
It is well known, and easy to check, that the above Serre relations, and
so the  Kac-Moody  algebra, are invariant under the  involution
which acts on the Chevalley generators as 
$$E_a \to \eta_a F_a,\  F_a \to \eta_a E_a,\ 
H_a\to -H_a
\eqno(2.13)$$
where $\eta_a=\mp 1$ for  any positive simple root. 
\par
The so called Cartan 
involution invariant subalgebra 
is that given by taking all minus signs. For any combination of signs
$\eta_a$ one can consider the corresponding invariant
generators which generate a sub-algebra. 
\par
When $\det A_{ab} > 0$ the Kac-Moody algebra is one of the finite
dimensional semi-simple Lie algebras classified by Cartan. When 
$\det A_{ab} = 0$ we find the well known affine Lie algebras. 
However, when  $\det A_{ab} < 0$ very little is known about these
algebras.  Indeed, apart from  a few  exceptional cases,  an explicit
formulation of the generators, or even their number, is known.
\par
As we explained in the introduction the Kac-Moody algebras that are
of most interest are the very extended Kac-Moody algebras [13]. 
Given any finite-dimensional simple Lie algebra  
${\cal G}$,  there is a well-known procedure for   constructing  a  
corresponding affine algebra ${\cal G^{+}}$ by adding a node  to the
Dynkin diagram   in a certain
way which is related to the properties of the highest root of ${\cal  
G^{}}$. One
may also further increase by one the rank of the algebra ${\cal G^{+}}$  
  by
adding  to the Dynkin diagram a further node that is attached to the   
affine node
by a single  line
[21]. This is called the overextension  ${\cal G^{++}}$.  
The very extension, denoted ${\cal G^{+++}}$, is found by adding yet
another   node to the Dynkin diagram that is attached to the overextended
node by one line [13].  The rank  of very extended exceeds
by three the rank  of the
finite-dimensional simple Lie algebra ${\cal G^{}}$ from which one
started.  
\par 
The $E_{11}$, or $E_8^{+++}$,  algebra contains the  
generators  $K^a{}_b$ at level 0,
corresponding to the $A_{10}$ subalgebra,  and the generators 
$$ R^{a_1a_2a_3}, R^{a_1a_2\dots a_6},R^{a_1a_2\ldots a_8,b}
\eqno(2.14)$$
 at levels zero, 1, 2 and 3 respectively [4].  
The generators of $E_{11}$ at higher levels are listed in references
[17, 11]. 
\par 
The corresponding Borel sub-algebra up to, and including,
level 3 obeys the commutation relations [4]
$$
[K^a{}_b,K^c{}_d]=\delta _b^c K^a{}_d - \delta _d^a K^c{}_b,  
\eqno(2.16)$$
$$  [K^a{}_b, R^{c_1\ldots c_6}]= 
\delta _b^{c_1}R^{ac_2\ldots c_6}+\dots, \  
 [K^a{}_b, R^{c_1\ldots c_3}]= \delta _b^{c_1}R^{a c_2 c_3}+\dots,
\eqno(2.17)$$
$$ [ K^a{}_b,  R^{c_1\ldots c_8, d} ]= 
(\delta ^{c_1}_b R^{a c_2\ldots c_8, d} +\cdots) + \delta _b^d
R^{c_1\ldots c_8, a} .
\eqno(2.18)$$
$$[ R^{c_1\ldots c_3}, R^{c_4\ldots c_6}]= 2 R^{c_1\ldots c_6},\ 
\ \ 
[R^{a_1\ldots a_6}, R^{b_1\ldots b_3}]
= 3  R^{a_1\ldots a_6 [b_1 b_2,b_3]}, 
\eqno(2.19)$$
where $+\ldots $ means the appropriate anti-symmetrisation. 
The above commutators can be deduced, using the Serre relations and   from
the identification of the Chevalley generators of $E_{11}$ which are given
by [4] 
$$E_a=K^a{}_{a+1}, a =1, \ldots 10, \ 
E_{11}= R^{91011},
\eqno(2.20)$$
and
$$ H_a= K^a{}_a-K^{a+1}{}_{a+1}, a=1,\ldots ,10, \ 
H_{11}=
-{1\over 3}(K^1{}_1+\ldots +K^8{}_8)+{2\over 3}(K^9{}_9
+K^{10}{}_{10}+K^{11}{}_{11}). 
\eqno(2.21)$$
The commutators involving the analogous    negative level generators
are given in [35]. 
\par
Non-linear realisation of a algebra ${\cal G}$ with respect to a
subalgebra 
${\cal H}$ is just a theory which is built from the group elements $g$ of
${\cal G}$ such that it is invariant under the  symmetry 
$$g\to g_0 g h
\eqno(2.22)$$
where $g_0$ are constant group elements and $h$ are depend on the same
variables as $g$. 
On of the advantages of a  non-linear realisation is that the dynamics is
largely specified by the algebra if the local subgroup ${\cal H}$ is
sufficiently large. 
In [4] the local subalgebra was chosen to be the  so called Cartan 
involution invariant subalgebra. Although one can use
this,  it  requires a  Wick rotation to get to
get a space-time with a signature with one minus sign. The
idea of  different signs was discussed in [10] and the
choice 
$\eta_1=+1$, all the remaining $\eta_a$'s all negative  was advocated in
[10,9] to get a space-time with a signature with one minus sign directly. 
However, as observed in equation (2.13), we find that one can take several
possible choices of signs for the
$\eta_a$'s resulting in theories with different space-time signatures.
This idea has also occurred to the author of the very recent paper   [19]
who has explored its consequences in  detail. 
 Since one is
interested in equations which are, after possible eliminations, second
order in derivatives, we believe that the dynamics is essentially
determined  if  any of the above  local subalgebra  is chosen. 
\par 
For the case of $E_{8}^{+++}$, the local sub-algebra was used in
reference [3,4] to express the  group element in the form 
$$
g  =exp(\sum_{a\le b}{\hat h_{a}{}^b K^a{}_b} ) exp {({A_{c_1\ldots c_3} 
R^{c_1\ldots c_3}\over 3!})exp( {A_{c_1\ldots c_6}
 R^{c_1\ldots c_6}\over 6!})} 
exp{({h_{c_1\ldots c_8,d} R^{c_1\ldots c_8,d}\over 8!})}
\ldots ,
\eqno(2.23)$$
where the fields  $\hat h^a{}_b$,
$A_{c_1\ldots c_3}$ and $A_{c_1\ldots c_6}$  depend on $x^\mu$. 
\par
The fields $\hat h^a{}_b$ encode the gravitational degrees of freedom and
are related to the vierbein  by the equation [3]
$$ e_\mu {}^a= (e^{\hat h})_\mu{}^a
\eqno(2.24)$$
From the non-linear
realisation of [3,4], we can identify  $A_{c_1\ldots c_3}$ 
 and $A_{c_1\ldots c_6}$  as  the
fields of the three form and its dual 
 of eleven dimensional supergravity. The field 
$h_{c_1\ldots c_8,d}$ plays the role of the dual field of gravity [4]. 
It is important for what follows to realise that the gauge fields that
occur in the group element are referred to the tangent space. 
 Given a solution it is then straightforward using the
above relations to construct the corresponding
$E_{11}$ group element.  Although one can choose whatever
parameterisation of the group element one likes, the brane solution will
take on a more universal form if we treat the off-diagonal components of
$h_a{}^b$ in the same way as the gauge fields and adopt the
parameterisation 
$$g=g_h g_A
\eqno(2.25)$$
where 
$$g_h  =exp(\sum _a h_{a}{}^a K^a{}_a ) 
exp(\sum _{a<b} h_{a}{}^b K^a{}_b)
\eqno(2.26)$$
and 
$$g_A=exp ({A_{c_1\ldots c_3} 
{R^{c_1\ldots c_3}\over 3!}}) exp( A_{c_1\ldots c_6}
 {R^{c_1\ldots c_6}\over 6!}) 
exp(h_{c_1\ldots c_8,d} {R^{c_1\ldots c_8,d}\over 8!})
\ldots ,
\eqno(2.27)$$
The vierbein  is then given by 
$$ e_\mu {}^a= e^{h_a{}^a}(e^ {\tilde h})_\mu{}^a
\eqno(2.28)$$
where ${\tilde h}_a{}^b=h_a{}^b-h_a{}^a\delta_a{}^b$ is the off-diagonal
part of $h_a{}^b$. This new parameterisation will only affect solutions
with an off diagonal component to the metric such as the
pp-wave. 
\par 
Although we have reviewed the construction of the non-linear
realisation for
$E_{8}^{+++}$ it is straightforward to generalise to any ${\cal
G}^{+++}$ once one has identified the preferred $SL(D)$ sub-algebra
associated with  gravity. 
\medskip
{\bf {3.  The half BPS solutions of eleven dimensional supergravity}}
\medskip
In this section we will cast  half BPS solutions of eleven dimensional
supergravity as $E_{11}$ group elements. We  show that they have a very
special form given in equation (1.1).  We begin by  giving  a detailed
discussion of the case of the M2 brane. 
\medskip
{\bf 3.1 The M2 brane}
\medskip
This solution [22] has a 
line element   given by 
$$ds^2=N_2^{-{2\over 3}}(-(dx_1)^2+(dx_2)^2 +(dx_{3})^2)
+N_2^{{1\over 3}}((dx_{4})^2+\ldots +(dx_{11})^2),
\eqno(3.1)$$
together with  a four form   field strength 
$$F_{123i}=\pm{\partial \over \partial x^i} N_2^{-1},\ i=4,\ldots ,11.
\eqno(3.2)$$
In these equations  $N_2$ is a harmonic function of the form 
$N_2=1+{k\over
r^6}$  where $r^2=(x_{4})^2+\dots +(x_{11})^2$
\par
Using equation (2.28),  we conclude that 
$$(e^h)_1{}^1=(e^h)_2{}^2=(e^h)_3{}^3=N_2^{-{1\over 3}},\ 
(e^h)_4{}^4=\ldots =(e^h)_{11}{}^{11}=N_2^{{1\over 6}}
\eqno(3.3)$$
while, taking the upper sign, the three form gauge field is given by 
$$ A_{123}= N_2^{-{1}}-1,\  A_{123}^T= 1-N_2.
\eqno(3.4)$$
 In this equation the superscript T makes it clear
that we are dealing with the gauge field expressed with respect to the 
tangent space. We do not use this symbol if the nature of the indices
already makes it clear we are dealing with a quantity referred to the
tangent space, i.e.  $A_{a_1a_2a_3}$. 
\par
Substituting these values into the $E_{8}^{+++}$ group element of equation
(2.25)  we find that the M2 brane corresponds to 
$$g= exp\bigg(-{1\over 2}ln N_2({2\over 3}(K^1{}_1+K^{2}{}_{2}+K^{3}{}_{3})
-{1\over 3}(K^4{}_4+\ldots K^{11}{}_{11}))\bigg)
exp\bigg((1-N_2)R^{123}\bigg)
\eqno(3.5)$$
Examining equations (2.20) and  (2.21) we recognise that the first factor
contains
$H_{11}$  and the second factor $E_{11}$ if we were to shift the indices
by $+8$ mod 8. We  can change $E_{11}=R^{91011} $ to 
$R^{123}$ by taking the commutator of the former with 
the following elements 
$$K^1{}_2,\ 2K^2{}_3,\ 3K^3{}_4,\ \ldots,\ 3K^8{}_9,\ 2K^9{}_{10},\ 
K^{10}{}_{11}
\eqno(3.6)$$
in an appropriate order. 
Using equation (2.20) we find that the  root corresponding to $R^{123}$ 
is  given by 
$$\beta_{M2}= \alpha_{11}+\alpha_{1}+2\alpha_{2}+3\alpha_{3}+\dots 
+3\alpha_{8}+2\alpha_{9}+\alpha_{10}
\eqno(3.7)$$
The Cartan sub-algebra element in $E_{8}^{+++}$ corresponding to 
 $R^{123}$ is 
$$\beta_{M2}^i H_i=H_{11}+H_{1}+2H_{2}+3H_{3}+\dots 
+3H_{9}+2H_{9}+H_{10}
$$
$$={2\over 3}(K^1{}_1+K^{2}{}_{2}+
K^{3}{}_{3})
-{1\over 3}(K^4{}_4+\ldots +K^{11}{}_{11})
\eqno(3.8)$$
Hence, we find that the M2 brane corresponds to the $E_{8}^{+++}$ group 
element 
$$g=\exp(-{1\over 2} \beta_{M2}\cdot H\ln N_2)\exp ((1-N_2)E_{\beta_{M2}})
\eqno(3.9)$$
\medskip
{\bf 3.2 The M5 brane}
\medskip
This solution [23] has the  line element   
$$ds^2=N_5^{-{1\over 3}}(-(dx_1)^2+(dx_2)^2+\ldots +(dx_{6})^2)
+N_5^{{2\over 3}}((dx_{7})^2+\ldots +(dx_{11})^2). 
\eqno(3.10)$$
where $N_5=1+{k\over r^3}$ and $r^2=(x_{7})^2+\ldots +(x_{11})^2$. While
the four form field strength is given by 
$$F_{ijklm}=\pm\epsilon_{ijlkm}{\partial\over \partial x^m} N_5
, \ i,j,k,l,m=7,\ldots ,11
\eqno(3.11)$$
From the the view point of this field strength the M5 is a magnetic 
brane, but in this paper we view all branes as electric branes and so
consider it as arising from the dual gauge field $A_{\mu_1\ldots \mu_6}$
whose dual field strength in tangent space is given by 
$$
F_{c_1\ldots c_7}=-{1\over 4 !} \epsilon_{c_1\ldots c_7b_1\ldots b_4}
F^{b_1\ldots b_4}
\eqno(3.12)$$
Taking the lower sign in equation (3.11),   we find that $F_{1\ldots
6 k}=\partial_kN_5^{-1}$ and so 
$$A^T_{1\ldots 6}=1-N_5
\eqno(3.13)$$
\par
The corresponding $E_{8}^{+++}$ group element is given by 
$$g=exp\bigg(-{1\over 2}ln N_5({1\over 3}(K^1{}_1+\ldots +K^6{}_6)
-{2\over 3}(K^7{}_7+\ldots +K^{11}{}_{11}))\bigg)exp((1-N_5)R^{1\ldots
6})
\eqno(3.14)$$
\par
We now wish to express the above group element in a more elegant form. 
Equation (2.19) expresses
$R^{6\ldots 11}$ as the commutator of 
$R^{678}$ with $E_{11}=R^{91011}$ and, taking into account the action of
$K^a{}_b$'s to change the former generator from $E_{11}$, we find the root
corresponding to $R^{6\ldots 11}$ is 
$2\alpha_{11}+\alpha_{6}+2\alpha_{7}+3\alpha_{8}+2\alpha_{9}+
\alpha_{10}$. However, the gauge field that occurs in the solution
corresponds to $R^{1\ldots 6}$ and its associated root $\beta_{M5}$ is found
to be 
$$\beta_{M5}=(2\alpha_{11}+\alpha_{6}+2\alpha_{7}+3\alpha_{8}+
2\alpha_{9}+\alpha_{10})$$
$$
+(\alpha_1+2\alpha_{2}+3\alpha_{3}+4\alpha_{4}+5\alpha_{5}+5\alpha_{6}+
4\alpha_{7}+3\alpha_{8}+2\alpha_{9}+\alpha_{10})
\eqno(3.15)$$
Using equation (2.21) we find that 
$$\beta_{M5}\cdot H={1\over 3}(K^1{}_1+\ldots +K^6{}_6)
-{2\over 3}(K^7{}_7+\ldots +K^{11}{}_{11})
\eqno(3.16)$$
It is then obvious that the group element corresponding to
the M5 brane of equation (3.13) can be written as 
$$g=\exp(-{1\over 2} \beta_{M5}\cdot H\ln N_5)\exp ((1-N_5)E_{\beta_{M5}})
\eqno(3.17)$$
\medskip
{\bf 3.3 The M Wave}
\medskip
The pp wave solution is given by [24] 
$$ds^2=-(1-K)(dx_1)^2+(1+K)(dx_2)^2-2Kdx_1dx_2+((dx_3)^2+\ldots
+(dx_{11})^2)
\eqno(3.18)$$
where $N_{pp}=1+K=1+{k\over r^7}$. The three and six form gauge fields
vanish and this solution is purely gravitational. The
non-trivial components of the vierbein are 
given by 
$$e_1{}^1={1\over \sqrt{1+K}},\ e_2{}^2={ \sqrt{1+K}},\ 
e_1{}^2=-{K\over \sqrt{1+K}},\ e_2{}^1=0
\eqno(3.19)$$
Using equation (2.28) we find that 
$$h_1{}^1=-{1\over 2}ln(1+K),\ h_2{}^2=+{1\over 2}ln(1+K),\ 
 h_1{}^2=-K
\eqno(3.20)$$
The associated group element of $E_8^{+++}$ is 
$$g=exp(-{1\over 2}ln N_{pp} \beta_{pp}\cdot H)exp(1-N_{pp})E_{\beta_{pp}}
\eqno(3.21)$$
since the roots associated to the generator $E_1=K^1{}_2$ is
$\alpha_1=\beta_{pp}$ and $\beta_{pp}\cdot H=K^1{}_1-K^2{}_2=H_1$.
\par
Hence,  the above half
BPS solutions of M theory when expressed in terms of $E_8^{+++}$ group
elements have a universal form which given by equation (1.1). 
\par 
 The M Monopole of M theory involves the imperfectly understood dual
graviton as the active field and we will treat it in the context of the
IIA theory below. 

\medskip
{\bf 4. Type II theories}
\medskip
It has been argued that both  IIA supergravity and IIB supergravity 
when suitably extended possess an $E_8^{+++}$ symmetry [4,5]. The
difference between the two theories arises from  the two different ways
the
$A_{9}$  sub-algebra associated with gravity, and so space-time, is
selected from
$E_8^{+++}$. Given the Dynkin diagram of $E_8^{+++}$ starting with the very
extended node one can find a 
$A_{9}$ sub-Dynkin diagram in only two ways. When we get to the
junction of $E_8^{+++}$ Dynkin diagram, situated at the node labeled 8, 
we can continue along the line with two further nodes taking only the
first node to belong to $A_9$ or we can find the final $A_9$ node by
taking it to be the only one in the other choice of direction at the
junction. These corresponding theories are IIA and IIB theories 
respectively.  
\par
In this section, we will consider  group elements of the form of 
the universal formula (1.1) and find what are the metric and
gauge fields. We will find that we recover all the well known half BPS
solutions of the type II theories including  the eight brane of the
massive IIA theory. 
\medskip
{\bf 4.1 IIA }
\medskip
The low level generators of  $E_8^{+++}$
when decomposed to the 
$A_{9}$ sub-algebra relevant to the IIA theory are given by [4] 
$$K^a{}_b, R^{a_1\ldots a_{q}} , q=0,1,\ldots ,8
\eqno(4.1)$$
In the non-linear realisation this implies a set of gauge fields which
is precisely in agreement with the field content of IIA supergravity. 
The NS-NS sector corresponds to $q-1=-1,1,5,7$ and the R-R sector to 
$q-1=0,2,4,6$. The $q=0$ generator, denoted $R$,  corresponding to the
dilaton. In addition to those listed in equation
(4.1) one also finds  a 
nine form generator which leads to nine form gauge field that is
associated with the massive IIA theory [16]. We will
consider this generator in section 4.3. The generators at higher levels
are tabulated in [16]. These include generators whose corresponding 
fields are the dual of gravity which will also lead to solutions as well
as an infinite number of further generators.  
\par
The generators of equation (4.1) obey the usual commutators with 
the $K^a{}_b$ generators of $A_{10}$ as well as 
$$ [R,R^{a_1\ldots a_q}]=c_q R^{a_1\ldots a_q} , \ 
[R^{a_1\ldots a_p},R^{a_1\ldots a_q}]= c_{p,q} 
R^{a_1\ldots a_{(p+q)}}
\eqno(4.2)$$
In these relations  
$$c_{p+1}=\eta {(p-3)\over 4}, \quad 
\eta=\cases{1,\quad R-R\cr
-1,\quad NS-NS\cr}
$$
and 
$$  c_{1,2}=-c_{2,3}=-c_{3,3}=c_{2,5}= c_{1,5}=2,\ 
c_{1,7}=3, \ c_{2,6}=2,\ c_{3,5}=1. 
\eqno(4.3)$$
Commutators involving higher level generators on the right-hand side than
those in equation (4.1) are not shown, but some of these are given later. 
\par
The $E_a$ Chevalley generators of $E_8^{+++}$  are given by [4] 
$$E_a=K^a{}_{a+1},\ a =1, \ldots, 9, \  E_{10}= R^{10},\  E_{11}=
R^{910};  
\eqno(4.4)$$
while the Cartan sub-algebra generators are 
$$ H_a= K^a{}_a-K^{a+1}{}_{a+1}, a=1,\ldots ,9,
\  H_{10}= -{1\over 8}(K^1{}_1+\ldots +K^9{}_9)+{7\over 8}K^{10}{}_{10}
-{3\over 2}R,
$$
$$  H_{11}= -{1\over 4}(K^1{}_1+\ldots +K^8{}_8)+{3\over
4}(K^9{}_9 +K^{10}{}_{10})+R. 
\eqno(4.5)$$
\par
The non-linear realisation of $E_8^{+++}$ with the appropriate
$A_9$ subalgebra relevant to IIA supergravity leads to at low levels to
the IIA supergravity action [3]. This calculation should be viewed
with  the hindsight provided by the relations to the
$E_{11}$ generators given in [4]. The group element was parameterised by 
$g = g_h g_A$
where $g_h$ was given in equation (2.28) with the indices restricted to be
ten or less and 
$$
   g_A = e^{(1/9!)A_{a_1 \cdots a_9}R^{a_1 \cdots a_9} }
      \, e^{(1/8!)A_{a_1 \cdots a_8}R^{a_1 \cdots a_8} }   
      \, e^{(1/7!)A_{a_1 \cdots a_7}R^{a_1 \cdots a_7} } 
 \times
$$
$$
   e^{(1/6!)A_{a_1 \cdots a_6}R^{a_1 \cdots a_6} }
   e^{(1/5!)A_{a_1 \cdots a_5}R^{a_1 \cdots a_5} } \, 
   e^{(1/3!)A_{a_1 a_2 a_3}R^{a_1 a_2 a_3} } \, 
   e^{(1/2!)A_{a_1 a_2}R^{a_1 a_2} } \, 
   e^{A_{a} R^{a}} \,
   e^{A R}\ldots .
\eqno(4.6)
$$
With this parameterisation the fields which appear in the group element
are those found in the IIA supergravity theory as formulated  in [3]. 
We note that the dilaton occurs at the end of the expression. 
 \par
For each generator in equation (4.1) we wish to construct the group
element corresponding to equation (1.1) and then read off the
corresponding solution. The generators of equation (4.1) which
corresponding to the highest weight states of the 
$A_{9}$ representations are $R^{10-p\ldots 910}$. We  use the
commutators of equation (4.2) to construct these generators in terms of
multiple  commutators of the Chevalley generators  and we  then find
that the corresponding roots are given by 
$$\hat \beta_{p+1}=-{(p+1)\over 8}(K^{1}{}_{1}+\ldots+
K^{10-p-1}{}_{10-p-1}) +{(7-p)\over 8}(K^{10-p}{}_{10-p}+\ldots+
K^{10}_{}{10}) +b_p R
\eqno(4.7)$$
where 
$$b_p=\cases{\eta {(p-3)\over 2},\quad p\le 6\cr
0,\quad\quad \ \ \ \  p=7\cr}
\eqno(4.8)$$
We want to consider the root $\beta_{p+1}$ associated with the generator 
$R^{1\ldots p+1}=E_{\beta_{p+1}}$ which is a lowest weight representation of
$A_{9}$.  Taking account of the required commutators  with the 
$A_{9}$ generators we find that 
$$\beta_{p+1}\cdot H={(7-p)\over 8}(K^{1}{}_{1}+\ldots+ K^{p+1}{}_{p+1})
-{(p+1)\over 8}(K^{p+2}{}_{p+2}+\ldots+ K^{10}{}_{10})
+b_p R
\eqno(4.9)$$
We now consider the group element of equation (1.1); 
$$g= exp(-{1\over 2}ln N_p \beta_{p+1}\cdot H)exp((1-N_p)E_{\beta_{p+1}})
\eqno(4.10)$$
To read off the values of the fields that occur in the ten dimensional IIA
supergravity theory we must cast the group element in the form of
equation (4.6). In particular,  we must put the dilaton factor at the end
of the expression. The dilaton-gauge parts of the group element 
of equation (4.10) are of the
form 
$$exp(-{b_p\over 2}ln N_p R)exp((1-N_p)E_{\beta_{p+1}})
\eqno(4.11)$$
which is equal to 
$$exp(N_p^{-{b_p c_{p+1}\over 2}}(1-N_p)E_{\beta_{p+1}})
exp(-{b_p\over 2}ln N_p R)
\eqno(4.12)$$
\par
Hence, the solution corresponding to the root $\beta_{p+1}$ has a metric given
by 
$$
ds^2=N_p^{- {(7-p)\over 8}}(-(dx_1)^2+(dx_2)^2+\ldots +(dx_{p+1})^2)
+N_p^{{(p+1)\over 8}}((dx_{p+2})^2+\ldots +(dx_{10})^2),
\eqno(4.13)$$ 
a dilaton given by 
$$e^A=(N_p)^{-{b_p\over 2}}, 
\eqno(4.14)$$
and a gauge field, with world indices, is given by 
$$A_{1\ldots p+1}=N_p^{-1}-1. 
\eqno(4.15)$$
In this last expression one must  take  account of the change from
tangent to world indices. 
\par
Thus from equation (1.1) we recover the 
half BPS solutions for the F1 string and  the NS5 brane, both in the NS-NS
sector, as well as the $p=0,2,4,6$ D-branes in the R-R sector [40]. We do
take 
$p=-1$ as the corresponding generator is just the dilaton generator $R$
which is in the Cartan subalgebra and so  falls outside the scheme we
have considered. We find a solution for $p=7$, but it is just that for
Minkowski space-time with trivial dilaton.  

\medskip
{\bf 4.2 IIB }
\medskip
The low level generators of $E_8^{+++}$ decomposed with respect to
the  $A_9$ sub-algebra relevant to the IIB theory are given by [5] 
$$
 K^a{}_b,\, R_s, \,  R^{c_1 c_2}_s,\, R_2^{c_1 \ldots c_4},\, R^{c_1
       \ldots c_6}_s,\,  R^{c_1 \ldots c_8}_s,\qquad s = 1,\, 2
\eqno(4.16)
$$
where $s$ can take values 1 or 2 corresponding to the NS-NS and R-R 
sectors respectively. The corresponding Goldstone fields are in agreement
with the field content of the IIB supergravity theory provided one adds
the appropriate dual fields. As for the case of the IIA theory we do not
add the dual of gravity although they are present and would lead to new
solutions. The higher order generators are
tabulated in [16]. 
\par
The  generators in equation (4.16) obey the usual relations with the $A_9$
generators $K^a{}_b$ as well as the relations 
$$
   [R^{c_1\cdots c_p}_{s_1}    ,   R^{c_1\cdots c_q}_{s_2}] = c_{p,q}^{s_1,
   s_2}R^{c_1\cdots c_{p+q}}_{s(s_1,s_2)}, 
$$
$$
   [R_1, R_s^{c_1\cdots c_q} ] = d_q^s R^{c_1\cdots c_q}_s,\quad [R_2,
   R_{s_1}^{c_1\cdots c_q} ] = \tilde d_q^s R^{c_1\cdots c_q}_{s(2,s_1)}.
\eqno(4.17)
$$
 In the last line, we have separately written the 
commutators  for the dilaton generator $R_1$ with
coefficient
$d_p^s$, and the axion generator $R_2$ with coefficient $\tilde d_p^s$.
The superscript $s = s(s_1,s_2)$ depends on the  fields in the
commutator and it satisfies the properties $s(1,1) = s(2,2) = 1, \quad
s(1,2) = s(2,1) = 2$.  The constants in the above commutation relations
are given by 
$$
   d_{p+1}^s = \eta^s{(p-3)\over 4},\quad \eta^1=-1, \eta^2=1
\eqno(4.18)
$$
and 
$$ 
   c_{2,2}^{1,2} = - c_{2,2}^{2,1} = -1 , \quad  
   c_{2,4}^{2,2}  = - c_{2,4}^{1,2}  = 4 , \quad
   c_{2,6}^{1,2}  = 1  ,\quad
   c_{2,6}^{1,1}  = - c_{2,6}^{2,2} = {1\over 2}
$$
$$ 
   \tilde d_2^1 = - \tilde d_{6}^{2} = - \tilde d_{8}^{2}  = 1, \quad
   \tilde d_2^2 = \tilde d_{6}^{1}  = \tilde d_8^1 = 0 
\eqno(4.19)
$$
Commutators involving generators at higher levels than those in
equation (4.16) on the right-hand side are not given, but are discussed
later on. 
\par
The Chevalley generators of $E_8^{+++}$,  as it appears in IIB, are given
by [5] 
$$
E_a=K^a{}_{a+1}, a=1,\ldots 8,\  E_9=R_1^{9 10},\ E_{10}=R_2,\ 
E_{11}=K^9{}_{10}.
\eqno(4.20)
$$
as well as 
$$
H_a=K^a{}_a -K^{a+1}{}_{a+1}, a=1,\ldots, 8,\
H_{9}=K^{9}{}_{9}+K^{10}{}_{10}+R_1-{1\over 4}\sum_{a=1}^{10}K^a{}_a, 
$$
$$
   H_{10}=-2R_1,\ H_{11}=K^{9}{}_{9}-K^{10}{}_{10}
\eqno(4.21)
$$ 
\par
The group element of $E_8^{+++}$  can
be  written, taking account of the local subalgebra,  as 
$$
   g =  g_h g_A, 
\eqno(4.22)$$
where $g_h$ is as in equation (2.28), but with indices restricted to be
ten or less and 
$$
   g_A = e^{(1/8!)A^2_{a_1\cdots a_8} R^{a_1\cdots a_8}_2} 
   e^{(1/8!)A^1_{a_1\cdots a_8} R^{a_1\cdots a_8}_1} 
   e^{(1/6!)(A^2_{a_1\cdots a_6}R^{a_1\cdots a_6}_2
   + A^1_{a_1\cdots a_6}R^{a_1\cdots a_6}_1)}
$$
$$
   \times  e^{(1/4!)A^2_{a_1\cdots a_4}R^{a_1\cdots a_4}_2} \,
   e^{(1/2!)(A^2_{a_1a_2}R^{a_1a_2}_{2} +
   A^1_{a_1a_2}R^{a_1a_2}_{1})} \, e^{A^2 R_2} \, e^{A^1 R_1}\dots .
\eqno(4.23)
$$ 
Carrying out the non-linear realisation of $E_8^{+++}$ at low levels one
finds [5] the IIB supergravity theory with the above parameterisation of
fields.  The  comparison with much of the known literature is 
facilitated by
 relabel $A^1=\sigma$ and $A^2=\chi$. 
\par
We now wish to calculate for each generator of the  $E_8^{+++}$ algebra
in equation (4.16) the corresponding  group element of the form of
equation (1.1) and then find what solution it corresponds to. Using the
above equations and following the same steps as for the IIA case, we
find  the roots
$\beta_{p+1}^s$ corresponding to the lowest weight $A_9$ states in the
representations of equation (4.16), i.e. 
$R_s^{12}, R_2^{1234},\ldots $.  The corresponding elements in
the Cartan sub-algebra are given by  
$$\beta_{p+1}^s\cdot H={(7-p)\over 8}(K^{1}{}_{1}+\ldots+ K^{p+1}{}_{p+1})
-{(p+1)\over 8}(K^{p+2}{}_{p+2}+\ldots+ K^{10}{}_{10})
+b_p^s R_1
\eqno(4.24)$$
where 
$$b_p^s=\cases{\eta^s {(p-3)\over 2},\quad {\rm otherwise}\cr
0,\quad p=7,s=1\cr}. 
\eqno(4.25)$$
\par
We now consider the group element of equation (1.1), namely 
$$g= exp(-{1\over 2}ln N_p^s \beta_{p+1}^s\cdot H)exp((1-N_p^s)E_{\beta_{p+1}^s})
\eqno(4.26)$$
To read off the values of the fields that occur in the ten dimensional IIB
supergravity theory we must reorder the dilaton and gauge parts of the
group element to bring it into the form of equation (4.23).  The solution
corresponding to the root
$\beta_{p+1}^s$ has a metric given by 
$$
ds^2=(N_p^s)^{- {(7-p)\over 8}}(-(dx_1)^2+(dx_2)^2+\ldots +(dx_{p+1})^2)
+(N_p^s)^{{(p+1)\over 8}}((dx_{p+2})^2+\ldots +(dx_{10})^2)
\eqno(4.27)$$ 
a dilaton given by 
$$e^A=(N_p^s)^{-\eta^s{(p-3)\over 4}}, 
\eqno(4.28)$$
and a gauge field given by 
$$A_{1\ldots p+1}^{T s}= (N_p^s)^{-1}-1
\eqno(4.29)$$
As before we have made the change to express the gauge field with
respect to the tangent space. 
\par
Equation (4.27-29) are just the half BPS solutions of the IIB 
supergravity theory. In particular the $p=1,5,7$ branes of the NS-NS
sector and the $p=-1,1,3,5,7$ branes of the R-R sector [40]. We note that
these include the instanton $p=-1$ solution of reference [27]. We also
find two  seven brane solutions, although the one in the NS-NS sector is
just Minkowski space-time with trivial dilaton and so its interpretation
as a seven brane is rather degenerate.  In section 5.2 we  will find
another seven brane solution corresponding to a higher level generator. 
\medskip
{\bf 4.3 Massive IIA }
\medskip
Remarkably, the low level generators of $E_8^{+++}$ when decomposed with
respect to $A_9$ subalgebra relevant to the IIA theory contain the
generators of equation (4.1) as well as the generator
$R^{a_1\ldots a_9}$ [16]. This corresponds in the
non-linear realisation to a rank nine anti-symmetric tensor gauge field 
$A_{a_1\ldots a_9}$. Such a  field has previously 
proved useful in reformulating the massive IIA theory  in such a way that
it has an eight brane solution [30,28]. This latter formulation can be
described  as a non-linear realisation [31] which involves the fields
corresponding to the generators of equation (4.1) as well as
$A_{a_1\ldots a_9}$.  In this theory, the nine form generator arises from
the commutator of lower rank generators as [31] 
$$
[ R^{a_1a_2}, R^{a_3\ldots a_9}]=-4 R^{a_1\ldots a_9}
\eqno(4.30)$$
This  relation also follows from  the $E_8^{+++}$ algebra as is
easily seen using the roots given in [16].  \footnote{$^1$}{The
underlying Kac-Moody algebra was not identified for the massive IIA
theory described as a non-linear realisation in [31], but if one excludes
the space-time translation generator then it is clear that it is also
$E_{8}^{+++}$.}  Using this equation and equation (4.7), we find that the
root
$\beta_9$  corresponding to  the $A_9$ lowest weight generator
$R^{12\ldots 9}$ gives rise to the Cartan sub-algebra element 
$$\beta_9\cdot H=-{1\over 8}(K^1{}_1+\ldots +K^9{}_9)-{9\over
8}K^{10}{}_{10} +{5\over 2} R
\eqno(4.31)$$
\par
The solution corresponding to the root $\beta_9$ has a
metric given by 
$$
ds^2=(N_8)^{{1\over 8}}((-dx_1)^2+(dx_2)^2+\ldots +(dx_{9})^2)
+(N_8)^{{9\over 8}}(dx_{10})^2),
\eqno(4.27)$$ 
  a dilaton given by 
$$e^A=(N_8)^{-{5\over 4}}, 
\eqno(4.28)$$
and a gauge field given by 
$$A_{1\ldots 9}= (N_8)^{-1}-1
\eqno(4.29)$$
This agrees with putting $p=8$ in the general formulae of equations
(3.13-15). It is precisely the eight brane solution found in reference
[30,28]. We will discuss the eleven dimensional origin of this solution
in the next section. 
\medskip
{\bf 5 Higher level branes }
\medskip
Clearly,  we can construct a group element of the form of equation (1.1)
for any root of the $E_8^{+++}$ algebra.  However,  we do not have an
explicit expression for the $E_8^{+++}$ non-linear realisation at higher
levels than that considered in the above sections and so we can not be
sure that such group elements will correspond to solutions of this theory.
Nonetheless, given the universal form of equation (1.1) 
 for the half BPS solutions it is encouraging
to think that this formula provides solutions in general. In this
section, we will find explicit expressions for the field configuration
corresponding equation (1.1) for  certain of  the higher order
generators of $E_8^{+++}$ in eleven dimensions.  We do this for the
eleven dimensional theory and the IIB theory 
\medskip
{\bf 5.1 Branes in  Eleven dimensions at level four}
\medskip
To find the roots associated with the branes at level four  it will 
first prove
useful to find the roots associated to the  generators  of $E_8^{+++}$ at
level three. The only such  generators are $R^{a_1\ldots a_8,b}$  and
they 
 arise [4] in terms of lower level generators
as the commutator 
$[R^{a_1\ldots a_6},R^{b_1\ldots b_3}]=3 R^{a_1\ldots a_6[b_1b_2,b_3]}$. 
The $A_{10}$ highest weight  is the generator  $R^{4\ldots
11,11}$ and  its associated root $\beta _{8,1}$ is  given by 
$\beta _{8,1}=\alpha_{11}+
(2\alpha_{11}+\alpha_{6}+2\alpha_{7}+3\alpha_{8}+2\alpha_{9}+
\alpha_{10})+\alpha_4+2\alpha_5+\ldots
+2\alpha_8+\alpha_9$.
The  corresponding Cartan subalgebra element is 
$\beta _{8,1}\cdot H=-(K^1{}_1+K^2{}_2+K^3{}_3)+K^{11}{}_{11}$. 
We note, using equation (1.1), that  the corresponding solution has the
correct metric to be Taub-Nut except for the off diagonal components
of the metric which  presumably arise from the dual gravity field. 
\par
At level four $E_8^{+++}$ contains the generators [17,11]
$$
R^a\ (1,2,3,4,5,6,7,8,5,2,4),\  R^{(ab)}_c,\ (0,1,2,3,4,5,6,7,4,1,4)
$$
and
$$  R^{a_1a_2a_3}_{b_1b_2}\ (0,0,1,2,3,4,5,6,4,2,4)
\eqno(5.1)$$ 
where the numbers in brackets are the positive integers $n_a$ for 
the corresponding root, i.e. $\alpha=\sum_a n_a \alpha_a$, of the $A_{10}$
highest weight components, $R^{11}, R^{11 11}_1$ and $R^{91011}_{12}$. 
The above generators satisfy the constraints 
$R^{(ab)}{}_b=0,\ R^{a_1a_2c}_{b_1c}=0$. 
Taking into account these conditions, we find that the $E_8^{+++}$ 
algebra at level four has the commutation relation 
\footnote{$^2$}{ We note in
passing that the one generator of SL(32) in the local subalgebra not
identified precisely in [35] is 
$R^a-R_a$}$$
[R^{a_1a_2a_3}, R^{b_1\ldots b_8,c} ]=
\epsilon^{b_1\ldots b_8 a_1a_2a_3} R^c-{1\over (D-1)}
\epsilon^{b_1\ldots b_8 [a_1a_2|c|}R^{a_3]}
+3\epsilon^{b_1\ldots b_8 [a_1a_2|e|}R^{a_3]c}{}_{e}
$$
$$+{3\over 2}\bigg( \epsilon^{b_1\ldots b_8 [a_1|ef|}R^{a_2 a_3]c}{}_{e f}
-{1\over (D-2)}\epsilon^{b_1\ldots b_8 c ef}R^{a_1a_2 a_3}{}_{e f}
\bigg)
$$
The highest weight components of the above generators  arise from
the following commutators 
$$[R^{123},R^{4\ldots 11,11}] \propto R^{11}+\dots,\
[R^{23 11},R^{4\ldots 11,11}] \propto R^{11 11}_1,\
[R^{3910},R^{4\ldots 11,11}] \propto R^{91011}_{12} 
\eqno(5.2)$$
 We
can now read off the roots corresponding to these level four generators. 
For example,
$R^{11}$ corresponds to the root 
$$\beta_1=\beta_{8,1}+\alpha_1+2\alpha_2+3\alpha_3+\ldots
+3\alpha_8+2\alpha_9+\alpha_{10}+\alpha_{11}
\eqno(5.3)$$
\par
We are interested in the electric branes and so we consider the lowest
weight generators $R^1$, $R^{1 1}_{11}$ and $R^{123}_{1011}$ with
corresponding roots 
$\beta_{1},\ \beta_{2,9}$ and $\beta_{3,7}$ respectively.
The  Cartan subalgebra generator associated with $R^1$  is  given by 
$$\beta_{1}\cdot H={2\over 3}K^{1}{}_{1}-{1\over 3} (K^2{}_2+\ldots
+K^{11}{}_{11})
\eqno(5.4)$$ 
and the corresponding solution, using equation (1.1), is 
$$
ds^2=(N_1)^{-{2\over 3}}(-(dx_1)^2)   
+(N_1)^{{1\over 3}}((dx_2)^2+\ldots +dx_{11})^2)
\eqno(5.5)$$ 
This has the form  of a D0 brane. 
\par
The electric brane corresponding to the generator $R^{(ab)}_c$ 
 has a Cartan subalgebra element  
$$
\beta_{2,9}\cdot H={5\over 3}K^{1}{}_{1}-{1\over 3} (K^2{}_2+\ldots
+K^{10}{}_{10})-{4\over 3}K^{11}{}_{11},
\eqno(5.6)$$
and the corresponding metric is 
$$
ds^2=(N_{2,9})^{-{5\over 3}}(-(dx_1)^2)   
+(N_{2,9})^{{1\over 3}}((dx_2)^2+\ldots +dx_{10})^2)+(N_{2,9})^{{4\over
3}}(dx_{11})^2 
\eqno(5.7)$$ 
Finally, the electric brane corresponding to the generator
$R^{a_1a_2a_3}_{b_1b_2}$ has a Cartan subalgebra element 
$$
\beta_{3,7}\cdot H={2\over
3}(K^{1}{}_{1}+K^{2}{}_{2}+K^{3}{}_{3})-{1\over 3} (K^{(4)}{}_4+\ldots
+K^{9}{}_{9})-{4\over 3}(K^{10}{}_{10}+K^{11}{}_{11}),
\eqno(5.8)$$
and the corresponding metric is 
$$
ds^2=(N_{3,7})^{-{2\over 3}}(-(dx_1)^2+(dx_2)^2+(dx_3)^2)   
+(N_{3,7})^{{1\over 3}}((dx_4)^2+\ldots +(dx_{9})^2)
$$
$$+(N_{3,7})^{{4\over
3}}((dx_{10})^2+(dx_{11})^2) 
\eqno(5.9)$$ 
\par 
As might be expected the potential solutions corresponding to $E_8^{+++}$ 
generators that have a more complicated structure than just a set of
anti-symmetric indices also have more complicated form. There is
an obvious correspondence between the indices on the generator and
the form of the solution. For example, $R^1$ has its world volume
in the 1 direction and the omitted indices are transverse and 
$R^{123}_{1011}$ has a "world volume" in the two separate parts in the 1,
2 and 3 directions and the 10 and 11 directions with the rest being
"transverse". This applies to all the branes in the previous sections
which have anti-symmetrised indices. Where symmetrised indices occur, such
as in equation (5.7) the situation seems a bit more complicated. 
\medskip
{\bf 5.2  Higher branes in IIB }
\medskip
In the IIB theory, the generators at the next levels beyond those of
equation (4.16) are  [16] 
$$  R^{a_1\ldots a_7,b}_1\  (0,0,0,1,2,3,4,5,4,2,2);\ S^{a_1\ldots
a_8}_2\  (0,0,1,2,3,4,5,6,4,3,3);
$$
$$\  R^{a_1\ldots a_{10}}_2\  (1,2,3,4,5,6,7,8,5,1,4)
\eqno(5.10)$$
Clearly, one could also write the last generator as an $A_{9}$ scalar.
These arise as the commutators 
$$
[ R^{a_1\ldots a_6}_1, R^{b_1b_2}_1]=-{1\over 2}R^{a_1\ldots
a_8}_1-R^{a_1\ldots a_6 [b_1,b_2]}_1,\  
[ R^{a_1\ldots a_6}_2, R^{b_1b_2}_2]={1\over
2}R^{a_1\ldots 
a_8}_1+R^{a_1\ldots a_6 [b_1,b_2]}_1,\
$$
$$[ R^{a_1\ldots a_4}_2, R^{b_1\ldots b_4}_2]=
-8 R^{a_1\ldots a_4 [b_1b_2b_3,b_4]}_1,\
$$
$$
[R^{a_1a_2}_2, R_1^{a_3\ldots
a_8}]=S^{a_1\ldots a_8}_2,\  [R^{a_1a_2}_1, R_2^{a_3\ldots a_{10}}]=
R^{a_1\ldots a_{10}}_2
\eqno(5.11)$$
The subscripts 1 and 2 correspond to the assignment to generalised R-R and
NS-NS sectors which are defined by the generalisation
of the rule for the commutators that is used for the supergravity fields,
namely the commutator of a R-R generator with a 
NS-NS generator gives a R-R generator and all other commutators give a
NS-NS generators. 
\par
The  generator $ R^{a_1\ldots a_7,b}_1$ leads to the  dual graviton
field. Its corresponding solution is an analogue of Taub-Nut and we
will not consider it further here. The generator $R^{1\ldots 10}_2$
leads to a rank ten gauge field which was conjectured in [16] to be the
gauge field associated with the space filling nine brane. 
 The root corresponding to the  generator
$R^{1\ldots 10}_2$ is 
$$\beta_{10}^2=\beta_8^2+\beta_2^1+K^1{}_1+K^2{}_2-K^{9}{}_{9}-K^{10}{}_{10}
=-(K^1{}_1+\ldots +K^{10}{}_{10})+3R_1
\eqno(5.12)$$
The corresponding solution is 
$$
ds^2=N(-(dx_1)^2+(dx_2)^2+\ldots +(dx_{10})^2),\ e^A=N^{-{3\over
2}}   
\eqno(5.13)$$ 
This has the form of a space filling nine brane and as there are no
transverse coordinates we expect that $N$ is a constant. This brane
belongs to the R-R sector and must be the space filling nine brane 
anticipated using world sheet arguments  in [29]. It would also be
interesting to make the connection between this work and the 
IIB supersymmetry algebra of reference [44] which incoorperates a ten form
field. 
\par
Finally, we find the solution corresponding to the lowest weight
generator $S_2^{1\ldots 8}$ whose corresponding root is given by 
$$\hat \beta^2_8=-(K^{9}{}_{9}+K^{10}{}_{10})-2R_1.
\eqno(5.14)$$
The metric part of the solution being 
$$
ds^2=-(dx_1)^2+(dx_2)^2\ldots (dx_8)^2+N((dx_9)^2+(dx_{10})^2),\  
\eqno(5.14)$$ 
which we recognise as a seven brane. In addition to the solution given in
section 4.2 we now have two seven brane solutions in the R-R sector. 
The one in this latter section is associated with the dual of the axion
and should be related to that given in reference [27,28]. However, more
seven branes were found in reference [42] and it would be interesting to
establish a precise comparison. 


\medskip
{\bf 6 Relations between the eleven dimensional, IIA and IIB theories}
\medskip
The fact that the eleven dimensional theory, the IIA and IIB theories
all have an underlying non-linear $E_8^{+++}$ symmetry is consistent with
the general belief that they are all limits of some underlying theory
often referred to as M theory. In a
non-linear realisation there is a one to one   correspondence 
between the  fields and the generators in the algebra outside the local
subgroup. Hence, given a field in for example eleven dimensional theory we
can identify its  $E_8^{+++}$ root and find in  the IIA or IIB theory the
field which corresponds to that root. As such,  we find a correspondence
between the fields in the eleven dimensional theory and those in the  IIA
and IIB theories. In this section, we will give these correspondences at
low levels and in particular use it to find how the  massive IIA theory
and its eight brane are related to the eleven dimensional theory.  
\par 
As such, viewing M theory as having a $E_8^{+++}$ symmetry allows a more
precise understanding of what M theory is. Indeed, it is clear from
the $E_8^{+++}$ perspective that the underlying theory does not have a
specified dimension.   An established piece of M theory
dogma is that it is a theory in eleven dimensions and so to distinguish
the approaches one might   call  the  underlying theory the $E_8^{+++}$
theory, or E-theory for short.  
\medskip
{ \bf {6.1 Correspondence between the eleven dimensional and IIA
theories}}
\medskip
The correspondence between the generators of the eleven dimensional 
formulation of $E_8^{+++}$  and the $E_8^{+++}$ formulation appropriate 
 to the IIA theory in  ten dimensions was given at low
levels in [4].  Demanding equivalence of
the Cartan subalgebra generators of the IIA theory in equations (4.5) and
the eleven dimensional theory in equation (2.21) one finds the relations
$$ \tilde K^a{}_a=K^a{}_a,\quad a=1\ldots 10,\quad 
\tilde R= {1\over 12}(-\sum _{a=1}^{10} K^a{}_a +8 K^{11}{}_{11}) . 
\eqno(6.1)$$
In this equation and all the equations in this subsection  we   treat
the eleventh index, denoted 11, as special while  the indices
$a,b,\ldots$ take the range 
$a,b,\ldots = 1,\ldots ,10$. We also 
denote the generators of the
IIA theory  with a $\tilde {\ }\ $. 
 Equating the simple root generators of equation 
(2.20) and equation (4.4) we find that 
$$ \tilde K^{a}{}_{a+1}=K^{a}{}_{a+1}, \quad a=1\ldots 9,\quad 
\tilde R^{10}=K^{10}{}_{11}, \quad
\tilde R^{910}=R^{91011}
\eqno(6.2)$$
Since by definition in a Kac-Moody algebra  all generators are formed
from the commutators of the simple root generators, the above equation 
fixes the correspondence for all generators. Comparing the resulting
commutators in the two theories we  find that  [4] 
$$ \tilde K^a{}_b= K^a{}_b,\ \tilde R^a=K^a{}_{11}, \ \tilde R^{a_1
a_2}=R^{a_1 a_2 11}, \ 
\tilde R^{a_1 a_2 a_3}=R^{a_1 a_2 a_3} $$
$$ \tilde R^{a_1 \ldots
a_5}=R^{a_1\ldots a_5 11}, \tilde R^{a_1 \ldots a_6}= -R^{a_1 \ldots
a_6},\ 
$$
$$
\tilde R^{a_1 \ldots a_7}= {1\over 2} R^{a_1 \ldots a_7 11,11}, 
\tilde R^{a_1 \ldots a_8}= {3\over 8}R^{a_1 \ldots a_8 ,11}  
\eqno(6.3)$$
\par
The relations between the fields can be  trivially read off from the
relations between the generators, for example $\tilde A_a=h_a{}^{11},\
\tilde A_{a_1 a_2}=A_{a_1 a_2 11}, \ 
\tilde A_{a_1 a_2 a_3}=A_{a_1 a_2 a_3} $ etc. 
These are of course consistent  at low levels with the
relations that relate the IIA theory to the eleven dimensional
supergravity theory by dimensional reduction on a circle [33]. 
However, we would stress that we are not regarding the eleven dimensional
theory as dimensionally reduced on a circle.  Substituting these
replacements into the eleven dimensional group element of equation (2.23) 
we would find, after a suitable rearrangement, the IIA group element of
equation (4.6). The rearrangement just leads to a set of field
redefinitions. 
\par
Some higher level generators and their corresponding roots are given in
[17,11] and [16] for the eleven dimensional and IIA theories respectively.
In particular, examining these tables one finds that 
the nine form generator
$R^{a_1\ldots a_9}$ of the 
$E_8^{+++}$ that arises in the IIA theory has the root
$(0,1,2,3,4,5,6,7,4,1,4)$ which  corresponds to the highest weight
component $R^{3\ldots 11}$. On the other hand, in  
the non-linear realisation of $E_8^{+++}$ with the $A_{10}$ subalgebra
that leads to the eleven dimensional theory the root 
$(0,1,2,3,4,5,6,7,4,1,4)$ is the highest weight component of the
level four generator  $R^{(ab)}_c$, that is  $R^{(1111)}{}_1$ or 
raising the lowered index with epsilon 
 $R^{(1111)}{}^{2\ldots 11}$. Since the $A_{9}$ subalgebra of $A_{10}$ is
in common we can identify all the components of the IIA nine form
generator from their highest weight components in the obvious way.
Performing such identifications for other generators we find that 
$$
\tilde R^{a_1 \ldots a_9}=\hat R^{(1111)}{}^{a_1 \ldots a_911},\ 
\tilde R^{a_1\ldots a_7,b}=R^{a_1 \ldots a_711,b},\ 
\tilde R^{b,a_1\ldots a_9}=R^{(b11)a_1\ldots a_911},\ 
$$
$$
\tilde R^{a_1a_2}_{b_1b_2}=\tilde R^{a_1a_2 11}_{b_1b_2},\ \tilde
R=R^{11},\ldots
\eqno(6.4)$$
 The precise 
coefficients can not be deduced by identifying  the generators 
 from their roots and are not necessarily one
as shown. They can be determined by comparison of the commutators in the
two theories as will be done elsewhere. 
\par
We can now trace the eleven dimensional origin of 
 the eight brane of the
massive IIA theory,  considered in section (4.3). 
As explained above, the  IIA generator $R^{a_1\ldots
a_9}$, which is associated with the  massive IIA theory, corresponds 
in the eleven dimensional theory to
the  level four generators $R^{(ab)}_c$, or equivalently
$R^{(ab) c_1\ldots c_{10}}$,  and so to the eleven dimensional field 
$A_{(ab)}^c$. This  strongly  suggests
that the eleven dimensional theory contains  not only the ten  dimensional
IIA supergravity [33], but also the massive IIA  supergravity theory [43]
when suitably truncated.
 However, to see this  one must
include in the non-linear realisation the fields corresponding to the
generators of  level four. This theory will be an extension of eleven
dimensional supergravity that has so far not been constructed and so
it is not surprising that this relation between the two theories has  not
have been noticed so far. Clearly,  it would also be interesting to
establish this connection in detail and find the relation between the
solution of equation (5.6) and the eight brane solution of equation
(4.27). This differs from the eleven dimensional interpretation of the
eight brane of [39]. 
\par
The above discussion side stepped the issue of space-time. In [35] 
 space-time was introduced into the non-linear realisation 
by considering  the fundamental representation, denoted
$l_1$  associated with the very extended node and taking the  semi-direct
product with $E_8^{+++}$. 
However, as discussed in [35], the
$l_1$ representation not only introduces the usual space-time coordinates,
but central charge coordinates  and an infinite number of other
coordinates arising from generators at higher levels. These are listed for
low levels for the eleven dimensional theory in  [35,9]. 
\par
The correspondence between the generalised coordinates in the two theories
can be found in much the same way as above. The highest weight state in
the $l_1$ representation corresponds to the space-time generator $P_1$. 
However, this is the same in both theories. The identification of
higher order generators  in the two theories can
then be found by the action of the  $E_8^{+++}$ generators in each
theory together with a knowledge of their identification. We find that 
$$\tilde P_a=P_a,\ \tilde Z= [\tilde P_{10}, \tilde R^{10}]=
[P_{10},  K^{10}{}_{11}]=P_{11}, 
\eqno(6.5)$$
and similarly at higher levels. In fact, the generator $\tilde Z$ occurs
as a central charge in the IIA supersymmetry algebra.  We note that both
theories contain fields that depend on the same number of generalised
coordinates and so from  this view point neither theory is preferred. 
\medskip
{ \bf {6.2 Correspondence between the eleven dimensional and IIB 
theories}}
\medskip
It is well known that if one reduces the ten dimensional IIA and IIB
supergravities on a circle to nine dimensions the resulting
supergravities coincide and  the reduced IIA and IIB  string
theories are related by T duality. However, it is generally not expected
that there exist explicit  relations between  the IIA and IIB theories in
ten dimensions, that is  without any dimensional reduction.  
However, as explained at the beginning of this section, as both theories 
are based on 
$E_8^{+++}$, we can relate their generators and hence their
fields.  In this case, the algebras corresponding to their gravity sectors
only overlap on a $A_{8}$ subalgebra and so they have the index ranges
$a,b,c,\dots =1,2\ldots ,9$ in common. In what follows  in this
subsection $a,b,c,\ldots$ are assumed to take this range.   Equating the
Cartan subalgebra elements in equation (2.21) and in equation (4.21) and
putting a
$\ \hat {}$ on all the IIB generators we find that 
$$\hat K^a{}{}_a=K^a{}{}_a,\quad a=1\ldots 9,\quad 
\hat K^{10}{}_{10}={1\over 3}\sum_{a=1}^9 K^{a}{}_{a}-{2\over
3}(K^{10}{}_{10}+K^{11}{}_{11})
$$
$$ 
\hat R_1=-{1\over
2}(K^{10}{}_{10}-K^{11}{}_{11})
\eqno(6.6)$$
\par
Identifying the  simple roots of equations (2.20) and (4.20) we find that 
$$
\hat K^{a}{}_{a+1}=K^{a}{}_{a+1}, \quad a=1\ldots 8,\quad 
\hat K^{9}{}_{10}=R^{91011},\quad \hat R_2=K^{10}{}_{11}, \quad
\hat R_1^{910}=K^{9}{}_{10}
\eqno(6.7)$$
This completely fixes the identification of the generators in the
two theories. To find the higher level identifications we use these
relations  in conjunction with the  commutators in the two theories. 
One finds that 
$$\hat R^{a10}_1=K^a{}_{10}, \ \hat R^{ab}_1=R^{ab11},\quad
\hat R^{a10}_2=-K^a{}_{11}, \ \hat R^{ab}_2=-R^{ab10},\quad
$$
$$
\hat R^{a_1\ldots a_3 10}_2=-R^{a_1\ldots a_3}, \ 
\hat R^{a_1\ldots a_4}_2=2R^{a_1\ldots a_4 10 11}\quad
$$
$$\hat R^{a_1\ldots a_510}_2=-{1\over 2} R^{a_1\ldots a_511},\ 
\hat R^{a_1\ldots a_6}_2={1\over 2}R^{a_1\ldots a_61011,11},\quad
$$
$$
\hat R^{a_1\ldots a_510}_1={1\over 2} R^{a_1\ldots a_510},\ 
\hat R^{a_1\ldots a_6}_1=-{1\over 2} R^{a_1\ldots a_61011,10},\quad
$$
$$
\hat R^{a_1\ldots a_710}_2={1\over 2} R^{a_1\ldots a_711,11},\ 
\hat R^{a_1\ldots a_7 10}_1=-{1\over 2} (R^{a_1\ldots a_710,11}+
R^{a_1\ldots a_711,10}),\quad
$$
$$
\hat R^{a_1\ldots a_7 10,10}_1=R^{a_1\ldots a_6},\ 
\hat S^{a_1\ldots a_710}_2=-{1\over 2}\hat R^{a_1\ldots a_610,10} 
$$  
$$\hat R^{a_1\ldots a_7 10,a}=-{3\over 4}( R^{a_1\ldots a_7
10,11}- R^{a_1\ldots a_7 11,10})
$$
$$ 
\hat R^{a_1\ldots a_610,b}={1\over 4}( R^{a_1\ldots a_6 b 11,10}-
R^{a_1\ldots a_6 b 10,11})- R^{a_1\ldots a_6  10 11,b}
\eqno(6.8)$$
The coefficients are deduced by comparison of the commutators of the two
theories. We have omitted  field comparisons that involve level four
generators in the eleven dimensional theory, but it is interesting to
observe that some generators  that are associated with the IIB
supergravity, including the dual fields, correspond to eleven dimensional
generators whose associated fields  are beyond those  found in the eleven
dimensional supergravity approximation. The field correspondence can be
read off i.e. 
$\hat A_{a10}^1=h_a{}^{10}, \ \hat A_{ab}^1=A_{ab11},\ldots$
 We note that the IIB ten 
form generator
$R^{a_1\ldots a_{10}}{}_2$ considered in equation (5.10), which gives
rise to the space filling nine brane, corresponds to the generator
$R^{(1111)}_{10}$, or equivalently 
$R^{(1111)}{}^{12\ldots 9 11}$, of the eleven
dimensional theory. Hence, the  eight brane of the massive IIA theory and
the space filling nine brane of the IIB theory have a common eleven
dimensional origin in different components of the field $A_{ab}^c$.
\par
We can also establish the correspondence between the generalised
coordinates as for the IIA case above. Again the highest weight states 
of the two $l_1$ representations corresponding to the two 
space-time generators $P_1$ can be identified. The
identification for all the other generators in the two $l_1$
representations then follows.  We find that 
$$\hat P_a=P_a,\ \hat P_{10}= [ \hat P_{9}, \hat K^9{}_{10}]=
[P_{9},  R^{91011}]=Z^{1011},\ \hat Z^{10}=[\hat P_9,\hat R_1^{910}]
=[P_9,K^9{}_{10}]=P_{10}
\eqno(6.9)$$
and similarly at higher levels. 
In the last step we used the relation $[P_{a}, 
R^{c_1c_2c_3}]=3\delta _a^{[c_1}Z^{c_2c_3]}$ [35]. Hence we must identify
the generalised coordinates as 
$$\hat x^a=x^a,\ \hat x^{10}=z_{1011},\  \hat z_{10}=x_{10},\ldots 
\eqno(6.10)$$
Thus  the equivalence of  the IIB theory with the eleven dimensional
theory requires an exchange of space-time coordinates with central
charges which is beyond the scope of the more usual
supergravity considerations.  We note that the two theories have a
dependence on the generalised coordintes that is in a one to one
cporrespondence and so  the IIB theory 
does not depend on less coordinates than the eleven dimensional theory. 
\medskip
{\bf 7 Discussion}
\medskip
In this paper we have concentrated on single brane solutions, but 
multi-brane solutions can be formed by taking superpositions of single
brane solutions [41]. In the  framework of this paper one can multiply two
group elements  and find a third which we may try to interpret as a
solution. Indeed,  given two branes whose corresponding group elements 
$g_1$ and $g_2$ are of the form of equation (1.1) with roots 
$\beta_1$ and $\beta_2$ their product is given by 
$$g_1 g_2= exp(-{1\over 2}ln N_1 \beta_1\cdot H)exp(1-N_1)E_{\beta_1}
exp(-{1\over 2}ln N_2\beta_2\cdot H)exp(1-N_2)E_{\beta_2}
$$
$$
=exp(-{1\over 2}ln N_1 \beta_1\cdot H-{1\over 2}ln N_2\beta_2\cdot H)
exp\{ (1-N_1)(N_2)^{{\beta_1\cdot \beta_2\over
2}}E_{\beta_1} \}exp(1-N_2)E_{\beta_2}
\eqno(7.1)$$
For example, one can consider the case of the M2 brane of equation (3.9)
with the M-wave of equation (3.21). In this case,
$\beta_1=\beta_{M2}$ and $\beta_2=\beta_{pp}$. One finds that 
$\beta_{M2}\cdot\beta_{pp}=0$
 and the resulting solution 
takes the form 
$$ds^2=N_2^{-{2\over
3}}(-(1-K)(dx_1)^2+(1+K)(dx_2)^2-2Kdx_1dx_2+((dx_3)^2) 
$$
$$+N_2^{{1\over
3}}((dx_{4})^2+\ldots +(dx_{11})^2).
\eqno(7.2)$$
The gauge field is given by $(N_2)^{-1}-1$. This  is indeed the  known 
solution for a M2 brane superimposed with a M wave. Similarly, we can 
find the known solution for two M2 branes in the  123 and 145 directions
by multiplying the corresponding group elements.  In this case, 
$\beta_1=\beta_{M2}$ and
$\beta_2=\beta_{M2}-(\alpha_2+2\alpha_3+\alpha_4)$. One finds these
roots are orthogonal and the solution derived from the product of the
group elements is again the known solution. One can also add yet another
M2 brane in the 167 direction and the group multiplication also gives the
known solution for three M2 branes. 
All these branes have
$1\over 4$  supersymmetry.  It would be interesting to  find the
$E_8^{+++}$ group element for the more general branes that preserve
$1\over 4$ supersymmetry and the general rules for constructing branes by
group multiplication of their corresponding group elements. 
\par
 In fact, there do exist other half
BPS solution than those considered in this paper, such as when the M2
brane lies within the M5 brane as considered in reference [36]. It would
be interesting to find the
$E_8^{+++}$ form of such solutions as well as those that preserve all 32
supersymmetries. The  elegant form of the group element for 
the usual half BPS
branes  leads one to think that  $E_8^{+++}$ may also be useful
for classifying solutions with a
given amount of supersymmetry. 
\par
The work in this paper also gives a new framework for considering 
the idea [37] that  the scattering of BPS branes defines some kind of
algebra. Given a BPS brane we can define its $E_8^{+++}$  group element
and so   the scattering of branes can be interpreted as an operation
on several copies of $E_8^{+++}$. An elementary example can be thought of
as  the above group multiplication to form composite branes. It would be
interesting to find what this operation is in general. Clearly, this is
related to the dynamics of the underlying non-linear realisation which we
have yet to bring into play. 
\par
In any  non-linear realisation the fields are encoded in a group
element and their transformations under the symmetries is given in
equation (2.22). In the more familiar non-linear realisations
used in particle physics, space-time is introduced in an adhoc way and
$g_0$ does not depend on space-time. As a result, a group transformation
does not change the space-time dependence of any solution. In the case
of the non-linear realisations considered here, the group element in one
 way or another involves space-time and so one can expect that
the possible group transformations that one can carry out on any solutions
are much more extensive. As such,  one can expect that $E_8^{+++}$
transformations will relate very large areas of the moduli space of
solutions. Certainly, the way the expression of solutions in terms of
group elements makes it particularly easy to carry out Weyl and other
transformations on the solutions. 
\par
The algebra $E_8^{+++}$ contains generators corresponding to the
supergravity fields and their dual. However,  in  this paper we have
regarded all branes as electric branes. As a result, when dealing with a
brane that is usually regarded as a magnetic brane, such as the M5 brane, 
we take the corresponding six rank dual gauge field to be the active
field.  Nonetheless, one might expect that the gauge fields and their
dual are related by a duality condition expressed through their field
strengths and so one can wonder  if both gauge fields should be active.
The resolution of this paradox probably relates to the fact that most 
brane solutions are not solutions of pure supergravity, but require
external sources which are also outside the non-linearly realised theory
considered here.  It is to be hoped that the incorporation of these
sources  allows  the purely electric  choice we have taken. 
\par
One could also consider  equation (1.1) for  generators none of whose
indices are in the 1, or time,  direction. The
resulting  potential brane solutions would have a world volume that  is
Euclidean and are likely to be  related to S branes. We will report on 
this possibility elsewhere.  It is straightforward to extend the
considerations in this paper to any 
${\cal G}^{+++}$ [38]. 
\par
In section six,  we explained that the common  $E_8^{+++}$ origin of
the   eleven dimensional 
theory and the IIA and IIB theories when viewed as 
non-linear realisations allows us to find explicit one to one
correspondences between any two of these  three theories. We did this
by relating the IIA and IIB theories to the eleven dimensional theory, but
this implies a similar correspondence between the IIA and IIB theories
which we could have found directly without involving the eleven
dimensional theory.  It is important to stress that this correspondence  
does not require a compactification to nine dimensions,  nor does the one
between the eleven dimensional and the IIA theory require a reduction on a
circle.  The correspondences allows us to find explicit relations between
the fields of the three theories. We also found that if we  regard
space-time to arise as part of the fundamental representation
$l_1$ associated with the very extended node, as advocated in [35], then
one can also find the correspondences between the
generalised coordinates of the three theories. This would also be the
case if one adopted the view point, as advocated as in [6] and [10], that
space-time is in some way contained in the  Kac-Moody algebra. We note
that it was observed [9]  that if this was the case then it was likely
that the  Kac-Moody algebra also encoded the central charge coordinates
and other generalised coordinates. We note that all three theories depend
on the same number of generalised coordinates which are in a one to one
correspondence. 
\par
Given the non-linear realisation for one theory, one can turn it into
the non-linear realisation of one of the other theories by carrying out
the changes of fields, and if needed the coordinates,   in the group
element and then rearranging it to be of the desired form. Thus, the three
theories seem to be much more closely related than previously thought. It
is important to note that  some of the dual fields associated with IIB
supergravity  and the nine form in the massive IIA theory
correspond to fields in the eleven dimensional theory that are at levels
higher than that which appear in the eleven dimensional supergravity
approximation. Also, the space-time coordinate 
$\hat x^{10}$ of the ten dimensional IIB theory becomes a component of
the central charge of the eleven dimensional theory as well  other
unusual changes. This is
beyond  the approximations of eleven dimensional supergravity [32] and the
IIA [33] and IIB supergravity [34] theories that have 
formed the basis for so much of our knowledge over recent years and it 
also involves effects not found in the context of string theory alone.  
\par
At the very least, it is encouraging to see that the
$E_8^{+++}$ algebra  contains generators leading to the eight brane of
the massive IIA theory and the space filling nine brane of the IIB theory
and we are able to find how these objects are related to the eleven
dimensional theory. 
\medskip
{\bf Acknowledgments}
I wish to thank Ulf Gran, Neil Lambert and George Papadopoulos for useful
discussions.  This research was supported in part by the PPARC grants 
PPA/G/O/2000/00451  and PPA/G/S4/1998/00613. 

\medskip

\medskip
{\bf References}
\medskip
\item{[1]} S. Ferrara, J. Scherk and B. Zumino, {\sl Algebraic
properties of extended supersymmetry}, Nucl. Phys. {\bf B 121} (1977)
393; E. Cremmer, J. Scherk and S. Ferrara, {\sl $SU(4)$ invariant
supergravity theory}, Phys. Lett. {\bf B 74} (1978) 61
\item{[2]} E. Cremmer and B. Julia, {\sl The $N=8$ supergravity
theory. I. The Lagrangian}, Phys. Lett. {\bf B 80} (1978) 48
\item{[3]} P.~C. West, {\sl Hidden superconformal symmetry in {M}
    theory },  JHEP {\bf 08} (2000) 007, {\tt hep-th/0005270}
\item{[4]} P. West, {\sl $E_{11}$ and M Theory}, Class. Quant.
Grav. {\bf 18 } (2001) 4443, {\tt hep-th/0104081}
\item{[5]} I. Schnakenburg and P. West, {\sl Kac-Moody Symmetries of
IIB supergravity}, Phys. Lett. {\bf B 517} (2001) 137-145, {\tt
hep-th/0107181} 
\item{[6]} T. Damour, M. Henneaux and H. Nicolai, {\sl $E_{10}$ and a
``small tension expansion'' of M-theory}, Phys. Rev. Lett. {\bf 89}
(2002) 221601, {\tt hep-th/0207267}
\item{[7]} T. Damour, M. Henneaux and H. Nicolai, {\sl Cosmological
billiards}, Class. Quant. Grav. {\bf 20} (2003) 020, {\tt
hep-th/0212256}
\item{[8]} E. Cremmer, B. Julia, H. Lu and C. N. Pope, {\sl Dualisation
of Dualities I}, Nucl. Phys. {\bf B 523} (1998) 73-144, {\tt
hep-th/9710119} 
\item{[9]} A. Kleinschmidt and  P. West, {\sl Representations of ${\cal
G}^{+++}$ and the role of space-time}, hep-th/0312247. 
\item{[10]} F. Englert and L. Houart, {\sl ${\cal G}^{+++}$ invariant
formulation of gravity and M-theories: Exact BPS solutions}, {\tt
hep-th/0311255} 
\item{[11]}  H. Nicolai and T. Fischbacher, {\sl Low Level
representations of $E_{10}$ and $E_{11}$}, Contribution to the
Proceedings of
the Ramanujan International Symposium on Kac-Moody Algebras and
Applications, ISKMAA-2002, Jan. 28--31, Chennai, India, {\tt
hep-th/0301017}
\item{[12]}  N. Lambert and  P. West, {\sl Coset symmetries in
dimensionally reduced bosonic string theory}, Nucl. Phys. {\bf B 615}  
(2001) 117, {\tt hep-th/0107209}
\item{[13]}  M. R. Gaberdiel, D. I. Olive and P. West, {\sl A
class of Lorentzian Kac-Moody algebras}, Nucl. Phys. {\bf B 645}
(2002) 403-437, {\tt hep-th/0205068}
\item{[14]}  F. Englert, L. Houart, A. Taormina and P. West,
{\sl The Symmetry of M-theories}, JHEP 0309 (2003) 020, {\tt hep-th/0304206}
\item{[15]} F. Englert, L. Houart and P. West, {\sl
  Intersection rules, dynamics and symmetries}, JHEP 0308 (2003) 025, 
{\tt hep-th/0307024}
\item{[16]} A. Kleinschmidt, I. Schnakenburg and P. West, {\sl
Very-extended Kac-Moody algebras and their interpretation at low
levels}, {\tt hep-th/0309198}
\item{[17]} P. West, {\sl Very-extended $E_8$ and $A_8$ at low
levels}, Class. Quant. Grav. {\bf 20} (2003) 2393,  hep-th/0307024. 
\item{[18]} V. Kac, {\sl Infinite dimensional algebras}, 3rd edition,
Cambridge University Press (1990)
\item{[19]} A. Keurentjes, {$E_{11}$ a sign of the times},
hep-th/0402090. 
\item{[20]} A. Miemiec and I.  Schnakenburg, {\it $E_{11}$ and
spheric vacuum solutions of eleven- and ten- dimensional supergravity
theories}; hep-th/0312096. 
\item{[21]} P. Goddard and D. Olive, {\it Algebras, lattices  
and strings,}
in {\it Vertex operators in Mathematics and Physics}, MSRI Publication  
no 3, Springer (1984) 51.
\item{[22]} M. Duff and K. Stelle, {\it Multimembrane solutions of
$d=11$  supergravity},  Phys. Lett. {\bf B253} (1991) 113. 
\item{[23]} R. Guven, {\it Black p-brane solutions of 11-dimensional
supergravity},  Phys. Lett. {\bf B276} (1992) 49. 
\item{[24]} M. Brinkman, Proc. Natl. Acad. Sci. USA 9 (1923) 1; C. Hull,
{\it Exact pp wave solutions of eleven dimensional supergravity}, Phys.
Lett. {\bf B139} (1984) 39. 
\item{[26]}S. Han and I. Koh, {\it $N=4$ supersymmetry remaining in
Kaluza-Klein monopole background in $D=11$ supergravity}, 
Phys. Rev. D31 (1985) 2503. 
\item{[27]} G. Gibbons, M. Green and M. Perry, {\it Instantons and
seven-branes in type IIB superstring theory}, hep-th/9511080. 
\item{[28]} E. Bergshoeff, M. de Roo, M. Green, G. Papadopoulos and P.
Townsend, {\it Duality of type II 7-branes and 8-branes}, hep-th/9601150. 
\item{[29]} J. Polchinski, Phys. Rev Lett. 75 (1995) 184. 
\item{[30]} E. Witten and J. Polchinski, Nucl. Phys. {\bf B438} (1995)
109. 
\item{[31]} I. Schnakenburg and P. West, {\it Massive IIA Supergravity
as a Non-linear realisation}, hep-th/0204207. 
\item{[32]} E. Cremmer, B. Julia and J. Scherk,  {\it  
Supergravity
theory in eleven dimensions}, Phys. Lett. {\bf 76B} (1978) 409. 
\item{[33]} C. Campbell and P. West, {\it $N=2$
$D=10$   non-chiral
supergravity and its spontaneous compactification}, Nucl.\ Phys.\ {\bf  
B243}
(1984) 112; M. Huq and M. Namazie, {\it Kaluza--Klein supergravity  
in ten
dimensions}, Class.\ Quant.\ Grav.\ {\bf 2} (1985) 293; 
F. Giani and M. Pernici, {\it $N=2$ supergravity in  
ten dimensions}, Phys.\ Rev.\ {\bf D30} (1984) 325.
\item{[34]} J. Schwarz and P. West, {\it Symmetries and
transformations of chiral
$N=2$ $D=10$ Supergravity}, Phys. Lett. {\bf 126B} (1983) 301; 
P. Howe
and P. West, {\it The Complete $N=2$ $D=10$ supergravity}, Nucl.\ Phys.\
{\bf B238} (1984) 181;J. Schwarz, {\it Covariant field equations of  
chiral $N=2$
$D=10$ supergravity}, Nucl.\ Phys.\ {\bf B226} (1983) 269.
\item{[35]} P. West, {\sl $E_{11}$, SL(32) and Central Charges},
Phys. Lett. {\bf B 575} (2003) 333-342, {\tt hep-th/0307098}
\item{[36]} J. Izquierdo, N. Lambert, G. Papadopoulos and P. Townsend, 
{\it Dyonic Membranes},  Nucl Phys {\bf  B460} (1996) 560,
hep-th/9508177. 
\item{[37]} J. Harvey and G. Moore, {\it Algebras BPS states, and
strings} 
hep-th/9510182, Nucl. Phys. {\bf B463} (1996) 315: {\it On the algebra of
BPS states}, hep-th/9609017. 
\item{[38]} P. Cook and P. West, shortly to be published. 
\item{[39]} N. Alonso-Alberca, P. Meessen and T. Ortin, {\it An SL(3, Z)
mutilpet of 8-dimensional type II supergravity theories and the gauged
supergravity inside }, hep-th/0012032.
\item{[40]} M. Duff and J. Lu, {Black and super p-branes in diverse
dimensions},  Nucl. Phys. {\bf B416} (1994) 301;
G. Gibbons and K. Maeda, {\it Black holes and membranes in higher
dimensional theories with dilaton fields}, 
 Nucl. Phys. {\bf B298} (1988) 741;
M. Duff and J. Lu, {\it The selfdual typeIIB superthreebrane}, 
Phys. Lett. {\bf 273B} (1991) 409; 
M. Duff and J. Lu, {\it Elementary five-brane solutions of $D=10$
supergravity},  Nucl. Phys. {\bf B354} (1991) 141; 
G. Horowitz and A. Strominger, {\it Black strings and Branes}, 
Nucl. Phys. {\bf B360} (1991) 197; 
C. Callan, J. Harvey and A. Strominger, {\it Supersymmetric string
solitons}, hep-th/9112030; 
A. Dabholkar, G. Gibbons, J. Harvey and F. Ruiz-Ruiz, {\it Superstrings
and solitons}, Nucl. Phys. {\bf B340} (1990) 33; 
\item{[41]} G. Papadopoulos and P. Townsend, {\it Intersecting M
branes}, Phys {\bf B380} (1996) 80;  A. Tseytlin, {\it Composite BPS
configurations of p-branes in 10 and 11 dimensions}, hep-th/9702163.
\item{[42]} E. Bergshoef, U. Gran and D. Roest,  {\it Type IIB seven brane
solutions from the nine-dimensional domain wall}, hep-th/0203202. 
\item{[43]} L. Romans, {\it Massive $N=2a$ supergravity in ten
dimensions},  Phys. Lett. {\bf B169} (1986) 374.
\item{[44]} E. Bergshoeff. M. de Roo, B. Janssen and  T. Ortin, {\it The
super D9-brane and its truncations}, hep-th/9901055. 
\end